\documentclass[12pt]{iopart}
\usepackage{amssymb}

\begin{document}

\title{Partial Observations, Einstein Locality and Bell Inequalities in
Quantized Detector Networks}
\author{George Jaroszkiewicz}
\address{School of Mathematical Sciences, University of Nottingham, \\
University Park, Nottingham NG7 2RD, UK}

\begin{abstract}
Quantized detector networks (QDN) deals with quantum information exchange
between observers and their apparatus rather than with systems under
observation. Partial observations in QDN involve subsets of the elementary
signal detectors which constitute an apparatus. We use them to prove that
QDN is consistent with Einstein locality and violations of Bell-type
inequalities.
\end{abstract}

\maketitle

\section{Introduction}

QDN (quantized detector networks) \cite{J2007C} is an approach to the
description of quantum experiments which emphasizes the role of observer and
apparatus, rather than the properties of any imagined SUO (system under
observation). QDN is based on the core principles underpinning Heisenberg's
approach to QM (quantum mechanics) \cite{HEISENBERG-1925, HEISENBERG-1927}
and asserts that the only physically relevant quantities in quantum physics
are signals from apparatus.

Non-locality has always been a fundamental issue in QM and is the source of
various apparent paradoxes, such as wave-particle duality and the
super-luminal transmission of certain types of information \cite%
{SCARANI-2000}. QDN interprets quantum non-locality as originating from the
fact that apparatus is invariably non-local, as are the processes of
extracting information from it, rather than reflecting strange,
non-classical properties of SUOs.\

There is a fundamental constraint on all quantum theories, known as \emph{%
Einstein locality}\textbf{\ }or the principle of local causes \cite%
{PERES:1993}. This principle asserts that \textquotedblleft \emph{events
occurring in a given spacetime region are independent of external parameters
that may be controlled, at the same moment, by agents located in distant
spacetime regions}\textquotedblright\ \cite{PERES:1993}. The aim of this
paper is to demonstrate that QDN indeed provides a consistent, physically
correct account of quantum physics, capable of satisfying Einstein locality
on the one hand and the demands of quantum non-locality, such as is seen in
violations of Bell-type inequalities, on the other.

The plan of this paper is as follows. In $\S 2$ we briefly review the core
formalism of QDN. In $\S 3$ we discuss labstates and maximal questions,
generalizing the latter in $\S 4$ to the notion of \emph{partial questions},
which is central to this paper. In $\S 5$ we discuss local operations on
apparatus and show how Einstein locality can be encoded into QDN. In $\S 6$,
we apply these ideas to a QDN discussion of local spatial rotations of
quantization axes in Stern-Gerlach experiments. This prepares the ground for
a discussion in $\S 7$ of EPR spin-pair experiments and Bell-type
inequalities. Finally, in $\S 8$, we discuss the implication of these ideas.

\section{QDN basics}

In QDN, time is measured in terms of quantum information exchange between
observer and apparatus and is discrete on that account. At any given time $n$%
, the observer's apparatus $\mathcal{A}_{n}$ consists of a finite number $%
r_{n}$ of ESDs (elementary signal detectors), each of which is represented
by a corresponding single signal qubit. In a classical approach, $\mathcal{A}%
_{n}$ would be represented by the Cartesian product of all the signal bits,
but in order to reflect quantum properties such as superposition and
entanglement, $\mathcal{A}_{n}$ is represented by a quantum register $%
\mathcal{R}_{n}\equiv \mathcal{Q}_{n}^{1}\otimes \mathcal{Q}_{n}^{2}\otimes
\ldots \mathcal{Q}_{n}^{r_{n}}$, the tensor product of all the signal qubits
at that time. Such a register, together with the contextual information as
to what each signal qubit means, is called a \emph{Heisenberg net}. Even if
the rank $r_{n}$ remains constant in time, an observer's Heisenberg net
changes at each time step: $\mathcal{R}_{n+1}$ is always distinct from $%
\mathcal{R}_{n}$.

In QDN, the observer calculates quantum outcome probabilities not via
quantum states of SUOs but via quantum states of their apparatus, which are
referred to as \emph{labstates.} A labstate at time $n$ is denoted by $|\Psi
,n)$ and is a normalized element of $\mathcal{R}_{n}$. Labstates can be
separable or entangled, but apparatus itself is not entangled normally
(although that possibility may exist but this cannot be discussed here).

In QDN, observations are answers to questions asked by observers of their
apparatus, and we shall use the two terms, \emph{observation} and \emph{%
question}, to mean the same thing. A \emph{maximal question} is one
involving all the ESDs in a given Heisenberg net, which invariably implies
that such an observation is a non-local operation. However, what is possible
in principle and in practice is that an observer could decide to look at
only a subset of the ESDs available to them at a given time, and then such
an observation is called a \emph{partial observation}.

In order to discuss partial observations, we first need to understand how
labstates are described. The two most useful representations of a current
labstate $|\Psi ,n)$ are in terms of the computational basis $\mathsf{B}%
_{n}\equiv \{|i,n):i=0,1,2,3,\ldots ,2^{r_{n}}-1\}$ and via the signal
operators $\left\{ \mathbb{A}_{i,n}^{+}:i=1,2,\ldots ,r_{n}\right\} $ and
their adjoints. The computational basis is useful for mathematical
calculations whilst the signal operators are directly tied in with the
intuitive physics of the situation.

Given a rank-$r$ quantum register $\mathcal{R}^{r}\equiv \mathcal{Q}%
^{1}\otimes \mathcal{Q}^{2}\otimes \ldots \otimes \mathcal{Q}^{r}$, there
are $r$ signal creation operators $\mathbb{A}_{i}^{+}:i=1,2,\ldots ,r$, each
of which has a corresponding signal destruction operator $\mathbb{A}%
_{i}:i=1,2,\ldots ,r$. These operators are constructed from tensor products
of various individual signal qubit operators, as discussed in detail in \cite%
{J2007C}.

In QDN, there is no concept of ground state. Instead, the nearest equivalent
to it is the\emph{\ void} state, or information vacuum, which represents an
apparatus in its quiescent state, i.e., one such that none of its
constituent ESDs would be in its signal (i.e., fired) state if examined by
the observer. The signal destruction operators annihilate the void state $%
|0) $, i.e., $\mathbb{A}_{i}|0)=0$, $i=1,2,\ldots ,r$ whilst the signal
creation operators create signal states, i.e., $\mathbb{A}%
_{i}^{+}|0)=|2^{i-1})$, $\mathbb{A}_{i}^{+}\mathbb{A}%
_{j}^{+}|0)=|2^{i-1}+2^{j-1})$, $i\neq j$, etc., using the computational
basis representation.

If $\mathbb{I}^{r}$ denotes the identity operator for $\mathcal{R}^{r}$ then
the signal operators satisfy the \emph{signal algebra}
\begin{eqnarray}
\{\mathbb{A}_{i},\mathbb{A}_{i}\} &=&\{\mathbb{A}_{i}^{+},\mathbb{A}%
_{i}^{+}\}=0,\ \ \ \{\mathbb{A}_{i},\mathbb{A}_{i}\}=\mathbb{I}^{r},\ \ \ \
\ i=1,2,\ldots ,r,  \nonumber \\
\lbrack \mathbb{A}_{i},\mathbb{A}_{j}] &=&[\mathbb{A}_{i},\mathbb{A}%
_{j}^{+}]=0,\ \ \ \ \ i\neq j,  \label{111}
\end{eqnarray}%
where square brackets denote commutators and curly brackets denote
anticommutators. We refer to the above as \emph{quadratic relations}, as
they involve products of two signal operators. The signal algebra (\ref{111}%
) is based on the physics of quantum observation, i.e., on what happens in
the laboratory, and is unique on that account.

It is convenient to define corresponding elementary projection operators
(EPOs). We define $\mathbb{P}_{i}\equiv \mathbb{A}_{i}^{+}\mathbb{A}_{i}$, $%
\overline{\mathbb{P}}_{i}\equiv \mathbb{A}_{i}\mathbb{A}_{i}^{+}$, $%
i=1,2,\ldots ,r$. These operators satisfy the quadratic relations%
\begin{equation}
\mathbb{P}_{i}+\overline{\mathbb{P}}_{i}=\mathbb{I}^{r},\ \ \ \mathbb{P}%
_{i}|0)=0,\ \ \ (0|\overline{\mathbb{P}}_{i}=0,\ \ \ \ \ i=1,2,\ldots ,r,
\end{equation}%
the \emph{cubic relations}%
\begin{eqnarray}
\mathbb{P}_{i}\mathbb{A}_{i} &=&\mathbb{A}_{i}^{+}\mathbb{P}_{i}=\overline{%
\mathbb{P}}_{i}\mathbb{A}_{i}^{+}=\mathbb{A}_{i}\overline{\mathbb{P}}_{i}=0,
\nonumber \\
\mathbb{P}_{i}\mathbb{A}_{i}^{+} &=&\mathbb{A}_{i}^{+}\overline{\mathbb{P}}%
_{i}=\mathbb{A}_{i}^{+},\ \ \ \mathbb{A}_{i}\mathbb{P}_{i}=\overline{\mathbb{%
P}}_{i}\mathbb{A}_{i}=\mathbb{A}_{i},\ \ \ i=1,2,\ldots ,r,
\end{eqnarray}%
and the \emph{quartic relations}%
\begin{eqnarray}
\mathbb{P}_{i}\mathbb{P}_{i} &=&\mathbb{P}_{i},\ \ \ \overline{\mathbb{P}}%
_{i}\overline{\mathbb{P}}_{i}=\overline{\mathbb{P}}_{i},\ \ \ \mathbb{P}_{i}%
\overline{\mathbb{P}}_{i}=\overline{\mathbb{P}}_{i}\mathbb{P}_{i}=0,\ \ \ \
\ i=1,2,\ldots ,r,  \nonumber \\
\lbrack \mathbb{P}_{i},\mathbb{P}_{j}] &=&[\mathbb{P}_{i},\overline{\mathbb{P%
}}_{j}]=[\overline{\mathbb{P}}_{i},\overline{\mathbb{P}}_{j}]=0,\ \ \ \ \
i\neq j.
\end{eqnarray}

\section{Labstates and maximal questions}

In this section, dependence on the temporal index $n$ is suppressed. Given a
rank-$r$ Heisenberg net, a pure labstate $|\Psi )$ is of the general form%
\begin{eqnarray}
|\Psi ) &=&\Psi _{0}|0)+\sum_{i=1}^{r}\Psi _{i}\mathbb{A}_{i}^{+}|0)+\sum_{1%
\leqslant i<j\leqslant r}\Psi _{ij}\mathbb{A}_{i}^{+}\mathbb{A}_{j}^{+}|0)+
\nonumber \\
&&\ \ \ \ \ \ \ \ \ \ \ \ \ \ \ \ \ \ \ \ \ \ \ \ \ \ \ \ \ \ \ \ \ \ \ \ \
\ \ \ \ldots +\Psi _{12\ldots r}\mathbb{A}_{1}^{+}\mathbb{A}_{2}^{+}\ldots
\mathbb{A}_{r}^{+}|0).
\end{eqnarray}%
Labstates are generally normalized to unity, so the coefficients satisfy the
condition%
\begin{equation}
(\Psi ,\Psi )=\left\vert \Psi _{0}\right\vert ^{2}+\sum_{i=1}^{r}|\Psi _{i}%
\mathbb{|}^{2}+\sum_{1\leqslant i<j\leqslant r}|\Psi _{ij}|^{2}+\ldots
+|\Psi _{12\ldots r}|^{2}=1.
\end{equation}%
For example, an arbitrary labstate for a rank-$2$ Heisenberg net is of the
form%
\begin{equation}
|\Psi )=\{\Psi _{0}+\Psi _{1}\mathbb{A}_{1}^{+}+\Psi _{2}\mathbb{A}%
_{2}^{+}+\Psi _{12}\mathbb{A}_{1}^{+}\mathbb{A}_{2}^{+}\}|0),  \label{123}
\end{equation}%
with $|\Psi _{0}|^{2}+|\Psi _{1}|^{2}+|\Psi _{2}|^{2}+|\Psi _{12}|^{2}=1$.
The interpretation of these coefficients is based on the Born rule in
standard quantum mechanics (SQM) \cite{BORN-1926}: if the apparatus is in
labstate (\ref{123}) prior to the observer looking at both ESDs
\textquotedblleft simultaneously\textquotedblright\ (which is possible in
QDN by definition), then the probability of each ESDs being found in its
void state is $|\Psi _{0}|^{2}$, the probability of ESD$_{1}$ being in its
fired state \emph{and} ESD$_{2}$ being in its void state is $|\Psi _{1}|^{2}$%
, the probability of ESD$_{1}$ being in its void state \emph{and} ESD$_{2}$
being in its fired state is $|\Psi _{2}|^{2}$, and the probability of both
ESDs being in their fired states is $|\Psi _{12}|^{2}$.

We introduce the following notation to encode the above ideas. Suppose the
observer looked at the $i^{th}$ ESD, $\mathcal{E}_{i}$, and obtained the
answer to the basic signal question \emph{What is the signal state of this
detector}? If the answer is \textquotedblleft void\textquotedblright , i.e.,
no signal, then we write $s_{i}=0$. Otherwise, if the answer is
\textquotedblleft fired\textquotedblright , i.e., there is a signal, then we
write $s_{i}=1$. What was expressed in words in the preceding paragraph can
now be expressed in terms of conditional probabilities. For example, $%
P(\{s_{1}=1\}\&\{s_{2}=0\}|\Psi )=|\Psi _{1}|^{2}$, and so on. A further
simplification is to express the various propositions symbolically. We write
$S_{i}\equiv \{s_{i}=1\}$\textbf{, }$\bar{S}_{i}\equiv \{s_{i}=0\}$ and
denote conjunctions such as $\{s_{i}=1\}\&\{s_{j}=0\}$ by $S_{i}\bar{S}_{j}$%
, etc. Then for example we write $P(\{s_{1}=0\}\ \&\ \{s_{2}=0\}|\Psi
)\equiv P\left( S_{1}S_{2}\right) =|\Psi _{0}|^{2}$, and so on. We may use
the properties of the projection operators given above to relate answers to
all these questions to expectation values of products of EPOs. For example,
for the rank-$2$ apparatus discussed above, we have four maximal questions
and answers, such as $P(S_{1}\bar{S}_{2}|\Psi )\equiv (\Psi |\mathbb{P}_{1}%
\overline{\mathbb{P}}_{2}|\Psi )=|\Psi _{1}|^{2}$, etc.

\section{Partial questions}

The above probabilities represent answers to maximal questions, i.e.,
questions which are asked of each and every ESD in the Heisenberg net at a
given time. For a rank-$r$ Heisenberg net, any maximal question involves a
product of $r$ distinct EPOs. For each ESD, $\mathcal{E}_{i}$, there are two
related EPOs, $\mathbb{P}_{i}$ and $\overline{\mathbb{P}}_{i}$, which form a
\emph{conjugate pair}. Therefore there are exactly $2^{r}$ distinct maximal
questions.

In the real world, however, observers could choose to ask \emph{partial
questions}, which involve looking at only some (or even none) of the ESDs.
An extreme example of a partial question is the normalization condition $%
(\Psi |\Psi )=1$. This is equivalent to asking for the probability of
finding anything at all, including no signals, without bothering to look.
This probability is obviously unity, conditional on the apparatus existing
in the first place and on a normalized labstate having been prepared.

It will be clear from the above that the set of all partial questions
involves expectation values of all possible products of the projection
operators. For the rank-$2$ example discussed above, there are four
non-trivial partial questions, such as $P(S_{1}|\Psi )\equiv (\Psi |\mathbb{P%
}_{1}|\Psi )=|\Psi _{1}|^{2}+|\Psi _{12}|^{2}$, etc. Trivial partial
questions are those for which the answer is always zero. For example, the
answer to the question \emph{What is the probability of ESD }$\mathcal{E}%
_{i} $ \emph{being in its void state }\textbf{and}\emph{\ in its signal
state? }is given by $P(\bar{S}_{i}S_{i}|\Psi )\equiv (\Psi |\overline{%
\mathbb{P}}_{i}\mathbb{P}_{i}|\Psi )=0$, which arises from the property that
$\overline{\mathbb{P}}_{i}\mathbb{P}_{i}=0$ for each $i$.

\section{Local operations}

In this section, we discuss a physical operation $U_{p}$ on a rank-$r$
apparatus $\mathcal{A}_{r}$ which affects a number $p$ of the ESDs in $%
\mathcal{A}_{r}$ and leaves the remaining $q\equiv r-p$ unaffected. The
affected ESDs and their corresponding signal qubits will be called \emph{%
local }whilst the unaffected ESDs and their corresponding signal qubits will
be called \emph{remote}.\ By unaffected, we mean that no possible partial
measurements on the remote ESDs alone would detect any changes, given that $%
U_{p}$ had been implemented\footnote{%
Note that the language here is imprecise. Experiments to detect changes in
the remote ESDs would actually involve ensembles of runs, comparing partial
measurements on apparatus evolving without the action of $U_{p}$ with
partial measurements on apparatus evolving with it.}.

The approach we take is to split the original register $\mathcal{R}^{r}$
into two sub-registers $\mathcal{R}^{p}$ and $\mathcal{R}^{q}$, such that $%
\mathcal{R}^{r}=\mathcal{R}^{p}\otimes \mathcal{R}^{q}$. $\mathcal{R}^{p}$
is the tensor product $\mathcal{Q}^{1}\otimes \mathcal{Q}^{2}\otimes \ldots
\otimes \mathcal{Q}^{p}$ of the local signal qubits\ whilst $\mathcal{R}^{q}$
is the tensor product $\mathcal{Q}^{p+1}\otimes \mathcal{Q}^{p+2}\otimes
\ldots \otimes \mathcal{Q}^{r}$ of the remote signal qubits. Each of these
subregisters comes with its own natural preferred basis $\mathsf{B}%
_{p}\equiv \left\{ |i)_{p}:i=0,1,\ldots ,2^{p}-1\right\} $ and $\mathsf{B}%
_{q}\equiv \{|a)_{q}:a=0,1,2,\ldots ,2^{q}-1\}$ respectively. We write $%
|i)_{p}\otimes |a)_{q}\equiv |i,a)$, an element of $\mathsf{B}_{r}$, the
natural basis for $\mathcal{R}^{r}$. Then orthonormality of the elements of $%
\mathsf{B}_{r}$ gives the rule $(i,a|j,b)=\delta _{ij}\delta _{ab}$.

The operation $U_{p}$ will be assumed here to leave the rank of the local
qubits unchanged, but in principle it is possible to consider changes in
rank. Such scenarios occur in particle decay experiments, for example, in
which case the rank increases monotonically with time \cite{J2007G}. It is
also possible to consider reduction in rank, such as occurs when apparatus
is destroyed, or when some ESDs are observed in order to transmit classical
information, such as occurs in teleportation experiments. In such cases, the
operators involved cannot be semi-unitary and non-linear quantum mechanics
is involved.

In our case, the action of $U_{p}$ will be represented by some semi-unitary
operator $\mathbb{U}_{p}$ acting on $\mathcal{R}^{r}$, taking it into a copy
$\mathcal{R}^{\prime r}$. Primes will denote objects such as ESDs, signal
qubits, EPOs and labstates \textbf{after} the action of $U_{p}$. To avoid
possible confusion, we write $|i,a)^{\prime }\equiv |\overline{i,a})$.

With these points in mind, then the most general local operation satisfying
these conditions has the following action on the natural basis elements of $%
\mathcal{R}^{r}:$%
\begin{equation}
\mathbb{U}_{p}|i,a)=\sum_{j=0}^{2^{p}-1}U_{p}^{ji}|\overline{j,a}),\ \ \ \ \
0\leqslant i<2^{p},\ \ \ 0\leqslant a<2^{q},
\end{equation}%
where the coefficients $\{U_{p}^{ji}\}$ are complex-valued functions of the
externally controlled parameters mentioned in the statement of the principle
of local causes in the introduction. These coefficients satisfy the
semi-unitarity relations%
\begin{equation}
\sum_{j=0}^{2^{p}-1}[U_{p}^{ji}]^{\ast }U_{p}^{jk}=\delta _{ik}.
\end{equation}%
From completeness of the basis set $\left\{ |i,a)\right\} $ we deduce%
\begin{equation}
\mathbb{U}_{p}=\sum_{i=0}^{2^{p}-1}\sum_{j=0}^{2^{p}-1}\sum_{a=0}^{2^{q}-1}|%
\overline{i,a})U_{p}^{ij}(j,a|.  \label{333}
\end{equation}

By inspection, it is clear that
\begin{eqnarray}
|\overline{j,a})(i,a|\mathbb{P}_{b} &=&\mathbb{P}_{b}^{\prime }|\overline{j,a%
})(i,a|,  \nonumber \\
|\overline{j,a})(i,a|\overline{\mathbb{P}}_{b} &=&\overline{\mathbb{P}}%
_{b}^{\prime }|\overline{j,a})(i,a|,\ \ \ \ \ p<b\leqslant r.
\end{eqnarray}%
Using this and the representation (\ref{333}), we readily find that the
operator $\mathbb{U}_{p}$ and the EPOs $\{\mathbb{P}_{a},\overline{\mathbb{P}%
}_{a}:p<a\leqslant r\}$ associated with the remote ESDs satisfy the relations%
\begin{equation}
\mathbb{U}_{p}\mathbb{P}_{a}=\mathbb{P}_{a}^{\prime }\mathbb{U}_{p},\ \ \
\mathbb{U}_{p}\overline{\mathbb{P}}_{a}=\overline{\mathbb{P}}_{a}^{\prime }%
\mathbb{U}_{p},\ \ \ \ \ p<a\leqslant r.
\end{equation}

To demonstrate that these are consistent with Einstein locality, consider an
actual experiment involving such a transformation $U_{p}.$ If $|\Psi )$ is
an initial labstate, i.e., before the action of $U_{p}$, then the final
labstate is $|\Psi ^{\prime })=\mathbb{U}_{p}|\Psi )$. Suppose the observer
performs arbitrary partial observations on the remote ESDs \emph{after} the
action of $U_{p}$. Then for any choice $\mathbb{P}_{a}\mathbb{P}_{b}\ldots
\overline{\mathbb{P}}_{z}$ of remote EPOs, we find%
\begin{eqnarray}
(\Psi ^{\prime }|\mathbb{P}_{a}^{\prime }\mathbb{P}_{b}^{\prime }\ldots
\overline{\mathbb{P}}_{z}^{\prime }|\Psi ^{\prime }) &=&(\Psi |\mathbb{U}%
_{p}^{+}\mathbb{P}_{a}^{\prime }\mathbb{P}_{b}^{\prime }\ldots \overline{%
\mathbb{P}}_{z}^{\prime }\mathbb{U}_{p}|\Psi )  \nonumber \\
&=&(\Psi |\mathbb{U}_{p}^{+}\mathbb{U}_{p}\mathbb{P}_{a}\mathbb{P}_{b}\ldots
\overline{\mathbb{P}}_{z}|\Psi )  \nonumber \\
&=&(\Psi |\mathbb{P}_{a}\mathbb{P}_{b}\ldots \overline{\mathbb{P}}_{z}|\Psi
),\ \ \ \ \ \ \ \ p<a,b,\ldots z\leqslant r,
\end{eqnarray}%
using the semi-unitarity condition $\mathbb{U}_{p}^{+}\mathbb{U}_{p}=\mathbb{%
I}_{r}$, the identity operator for $\mathcal{R}^{r}$. These probabilities
are obviously independent of the details of $U_{p}$, which proves that there
is no way that measurements on the remote ESDs alone could detect any
effects of the local operation $U_{p}$ acting on the local ESDs.\ This is
precisely what the principle of local causes requires.

A technical question remains about the evolution of the remote ESDs
themselves. The above analysis assumed that the remote ESDs evolved
unchanged during the action of $U_{p}$. This is equivalent to a \emph{null}
test on the remote ESDs, which would be unrealistic in practice. Therefore,
the discussion should be extended to two or more independent local
operations, one of which is the $U_{p}$ discussed above and the other is
some operation $V_{q}$ on the remote qubits. As before, basis elements for $%
\mathcal{R}^{r}$ are written in the form $|i,a)$, where $0\leqslant i<2^{p}$
and $0\leqslant a<2^{q}$, where $r=p+q$.

In the following we shall use the summation convention. Then the operator $%
\mathbb{U}_{p,q}$ for the combined simultaneous transformation is given by%
\begin{equation}
\mathbb{U}_{p,q}=|\overline{i,a})U_{p}^{ij}V_{q}^{ab}(j,b|,
\end{equation}%
where the indices $i$,$\ j$ are summed from $0$ to $2^{p}-1$ and $a,b$ are
summed from $0$ to $2^{q}-1$ and the coefficients satisfy the semi-unitarity
conditions
\begin{equation}
\lbrack U_{p}^{ij}]^{\ast }U_{p}^{ik}=\delta _{jk},\ \ \ [V_{q}^{ab}]^{\ast
}V_{q}^{ac}=\delta _{bc}.
\end{equation}%
Unlike the previous situation, all partial observations on either set of
localized signal qubits are now affected by the transformation. However, how
they are affected is still in accordance with Einstein locality, which is
proven as follows. Using the summation convention, suppose $|\Psi )=$ $\Psi
_{ia}|i,a)$ is an initial normalized labstate and consider a set of partial
observations on the first local set of ESDs represented by $(\Psi ^{\prime }|%
\mathbb{P}_{i_{1}}^{\prime }\mathbb{P}_{i_{2}}^{\prime }\ldots \overline{%
\mathbb{P}}_{i_{k}}^{\prime }|\Psi ^{\prime })$, where $1\leqslant
i_{1},i_{2},\ldots ,i_{k}\leqslant p$. Then%
\begin{equation}
(\Psi ^{\prime }|\mathbb{P}_{i_{1}}^{\prime }\mathbb{P}_{i_{2}}^{\prime
}\ldots \overline{\mathbb{P}}_{i_{k}}^{\prime }|\Psi ^{\prime
})=[U_{p}^{mn}]^{\ast }U_{p}^{ij}[V_{q}^{cd}]^{\ast }V_{q}^{ab}\Psi
_{nd}^{\ast }\Psi _{jb}(\overline{m,c}|\mathbb{P}_{i_{1}}^{\prime }\mathbb{P}%
_{i_{2}}^{\prime }\ldots \overline{\mathbb{P}}_{i_{k}}^{\prime }|\overline{%
i,a}).
\end{equation}%
By inspection, it can be seen that $(\overline{m,c}|\mathbb{P}%
_{i_{1}}^{\prime }\mathbb{P}_{i_{2}}^{\prime }\ldots \overline{\mathbb{P}}%
_{i_{k}}^{\prime }|\overline{i,a})=(\overline{m,0}|\mathbb{P}%
_{i_{1}}^{\prime }\mathbb{P}_{i_{2}}^{\prime }\ldots \overline{\mathbb{P}}%
_{i_{k}}^{\prime }|\overline{i,0})\delta _{ac}$, from which we deduce%
\begin{equation}
(\Psi ^{\prime }|\mathbb{P}_{i_{1}}^{\prime }\mathbb{P}_{i_{2}}^{\prime
}\ldots \overline{\mathbb{P}}_{i_{k}}^{\prime }|\Psi ^{\prime
})=[U_{p}^{mn}]^{\ast }U_{p}^{ij}\Psi _{nb}^{\ast }\Psi _{jb}(\overline{m,0}|%
\mathbb{P}_{i_{1}}^{\prime }\mathbb{P}_{i_{2}}^{\prime }\ldots \overline{%
\mathbb{P}}_{i_{k}}^{\prime }|\overline{i,0})  \label{444}
\end{equation}%
using the semi-unitarity conditions $[V_{q}^{ad}]^{\ast }V_{q}^{ab}=\delta
_{bd}$. The right hand side of (\ref{444}) is independent of any of the $%
V_{q}^{ab}$ coefficients parametrizing the $V_{q}$ transformation, which
proves that Einstein locality holds for $V_{q}$. The same argument applies
for $U_{p}$.

It is clear that this result generalizes immediately to apparatus of any
rank and to arbitrary splits involving arbitrary localized transformations,
provided none of these overlap as far as the ESDs involved are concerned. It
should be clear also that this formalism provides a basis for a discussion
of lightcone and causal set structure in a QDN approach to relativity (the
objective of future papers).

Two important conclusions can be drawn from this analysis: $i)$ it is
consistent to apply QM\ to parts of the universe, whilst ignoring the rest,
even though all of it is subject to the laws of quantum mechanics \cite%
{J2005A} and $ii)$ it is the possibility of isolating apparatus which gives
rise to the SUO concept. From the QDN perspective, it is not that particles
such as electrons and photons \textquotedblleft exist\textquotedblright ,
but that apparatus behaves in such as way as to support that notion, \emph{%
most of the time}.

\section{Local spatial rotations}

In this section we use the above results to prepare the ground for a
discussion of EPR spin pairs and Bell inequalities. Consider an experiment
involving an isolated Stern-Gerlach (S-G) apparatus $\Sigma \left( \mathbf{a}%
\right) $, where $\mathbf{a}$ is the associated quantization axis, together
with miscellaneous other equipment. $\mathcal{Q}^{1}$and $\mathcal{Q}^{2}$
are the two signal qubits associated with the two outcomes of $\Sigma (%
\mathbf{a})$ (known conventionally as \emph{spin-up} and \emph{spin-down}
respectively), and together constitute our local qubits. Signal qubits $%
\mathcal{Q}^{3},\mathcal{Q}^{4},\ldots ,\mathcal{Q}^{r},r>2,$ represent the
rest of the apparatus, which is considered remote.

Any S-G apparatus such as $\Sigma (\mathbf{a})$ is associated with a
definite axis of quantization $\mathbf{a}$ in physical three-space and this
axis can be altered by physically rotating the apparatus whilst doing
nothing to the rest of the laboratory. Suppose the initial axis is given by
the unit vector $\mathbf{k}$, which may be imagined to point in the
conventional $z$-direction.\ Unprimed quantities will be associated with
this orientation of the axis. Now consider a local operation $U(\mathbf{a})$
on $\Sigma \left( \mathbf{k}\right) $, rotating its axis $\mathbf{k}$ into
some new direction $\mathbf{a}$, such that $\Sigma (\mathbf{k})\rightarrow
\Sigma (\mathbf{a})$. Primed quantities will be associated with the new
orientation $\mathbf{a}$ of the axis. There are four kind of basis labstate
we need to consider, given by $|0,a)$, $\mathbb{A}_{1}^{+}|0,a)\equiv |1,a),%
\mathbb{A}_{2}^{+}|0,a)\equiv |2,a)$ and $\mathbb{A}_{1}^{+}\mathbb{A}%
_{2}^{+}|0,a)\equiv |3,a)$, where $0\leqslant a<2^{q}$ and $q=r-2$. These
are discussed in turn:

\begin{enumerate}
\item[$i)$] When isolated apparatus is in its void state, we would not
normally expect it to generate signals spontaneously whilst the apparatus is
being moved around in physical space. Hence we require rotations of S-G axes
of magnetization to satisfy the condition%
\begin{equation}
\mathbb{U}(\mathbf{a})|0,a)=|\overline{0,a}),\ \ \ \ \ 0\leqslant a<2^{q}.
\end{equation}

This supposes that space is homogeneous and isotropic. We expect this
condition to be broken in the presence of what would normally be regarded as
a gravitational field. This is analogous to the phenomenon of \emph{Rindler
radiation},\emph{\ }or the spontaneous creation of particles in accelerated
frames of reference,as discussed in conventional approaches to quantum
physics in the presence of curved spacetime.

\item[$ii)$] $|1,a)$ and $|2,a)$ are labstates representing the \emph{spin-up%
} and \emph{spin-down} outcomes of the S-G sub-experiment respectively,
relative to the current quantization axis. Given that $U(\mathbf{a})$ is an
active rotation of the quantization axis from $\mathbf{k}$ to $\mathbf{a}$,
then experience with the SQM description of the S-G experiment leads us to
write%
\begin{eqnarray}
\mathbb{U}(\mathbf{a})|1,a) &=&\alpha (\mathbf{a})|\overline{1,a})+\beta (%
\mathbf{b})|\overline{2,a}),  \nonumber \\
\mathbb{U}(\mathbf{a})|2,a) &=&\gamma (\mathbf{a})|\overline{1,a})+\delta (%
\mathbf{a})|\overline{2,a}),\ \ \ \ \ 0\leqslant a<2^{q},
\end{eqnarray}%
where the complex-valued coefficients $\alpha (\mathbf{a}),\beta (\mathbf{a}%
),\gamma (\mathbf{a})$ and $\delta (\mathbf{a})$ satisfy the semi-unitarity
conditions%
\begin{equation}
|\alpha (\mathbf{a})|^{2}+|\beta (\mathbf{a})|^{2}=|\gamma (\mathbf{a}%
)|^{2}+|\delta (\mathbf{a})|^{2}=1,\ \ \ \alpha (\mathbf{a})^{\ast }\gamma (%
\mathbf{a})=-\beta (\mathbf{a})^{\ast }\delta (\mathbf{a}).
\end{equation}

\item[$iii)$] Labstates of the form $|3,a)$ would not normally be
encountered in conventional S-G experiments, due to charge conservation.
Equivalently, a single photon entering a conventional beam-splitter would
not split into a photon pair. Hence we are entitled to assume$\ \mathbb{U}(%
\mathbf{a})|3,a)=|\overline{3,a})$, $0\leqslant a<2^{q}$, because any phase
can always be absorbed by a suitable redefinition of the outcome basis
elements.
\end{enumerate}

Together, these conditions give the representation%
\begin{eqnarray}
\mathbb{U}(\mathbf{a}) &=&\sum_{a=0}^{2^{q}-1}\left[ |\overline{0,a}%
)(0,a|+\{\alpha (\mathbf{a})|\overline{1,a})+\beta (\mathbf{a})|\overline{2,a%
})\}(1,a|\right.  \nonumber \\
&&\left. +\{\gamma (\mathbf{a})|\overline{1,a})+\delta (\mathbf{a})|%
\overline{2,a})\}(2,a|+|\overline{3,a})(3,a|\right]
\end{eqnarray}

\section{EPR spin-pair experiments}

We now extend the discussion to spin-pair experiments, which have been used
to explore issues in QM such as non-locality and violations of Bell-type
inequalities. Consider an apparatus consisting of many ESDs, four of which
are associated with a spin-zero bound state of two spin-half constituents,
such as an electron and a positron. Suppose a quantization axis $\mathbf{k}$
is chosen, and let $\mathcal{Q}^{1}$ and $\mathcal{Q}^{2}$ represent the two
spin polarization outcomes associated with constituent $\#1$, whilst $%
\mathcal{Q}^{3}$ and $\mathcal{Q}^{4}$ represent those for constituent $\#2$%
. It is traditional to describe such experiments in terms of a local
observer \emph{Alice} using S-G apparatus $\Sigma _{A}(\mathbf{k)}$ to
observe constituent $\#1$ whilst a remote observer \emph{Bob} uses S-G
apparatus $\Sigma _{B}(\mathbf{k})$ to observe constituent $\#2$. Other ESDs
in the apparatus are isolated from those used by Alice and Bob and are
represented by signal qubits $\mathcal{Q}^{5},\mathcal{Q}^{6}$, $\ldots ,%
\mathcal{Q}^{r}$, where $r$ is the current rank of the total apparatus.

Experience with spin-zero bound states in SQM leads us to take the initial
labstate $|\Psi )$ to have the form%
\begin{equation}
|\Psi )=\frac{1}{\sqrt{2}}\{\mathbb{A}_{1}^{+}\mathbb{A}_{4}^{+}-\mathbb{A}%
_{2}^{+}\mathbb{A}_{3}^{+}\}|\Phi ).
\end{equation}%
Here $|\Phi )$ is a normalized state in the total register $\mathcal{R}^{r}$
such that $\mathbb{P}_{1}|\Phi )=\mathbb{P}_{2}|\Phi )=\mathbb{P}_{3}|\Phi )=%
\mathbb{P}_{4}|\Phi )=0$. Equivalently, $\overline{\mathbb{P}}_{1}|\Phi )=%
\overline{\mathbb{P}}_{2}|\Phi )=\overline{\mathbb{P}}_{3}|\Phi )=\overline{%
\mathbb{P}}_{4}|\Phi )=|\Phi )$.

Before Alice and Bob perform any observations on their constituents, each
rotates the magnetization axis of their respective S-G apparatus
independently of the other. If Alice performs the rotation $\Sigma _{A}(%
\mathbf{k})\rightarrow \Sigma _{A}(\mathbf{a})$ and Bob performs the
rotation $\Sigma _{B}(\mathbf{k})\rightarrow \Sigma _{B}(\mathbf{b})$, where
$\mathbf{a}$ and $\mathbf{b}$ are unit three-vectors, then the operator $%
\mathbb{U}(\mathbf{a},\mathbf{b})$ representing the combined transformation
has the following action:%
\begin{eqnarray}
\mathbb{U}(\mathbf{a},\mathbf{b})\mathbb{A}_{1}^{+}\mathbb{A}_{4}^{+}|\Phi )
&=&\{\alpha (\mathbf{a})\mathbb{A}_{1}^{\prime +}+\beta (\mathbf{a})\mathbb{A%
}_{2}^{\prime +}\}\{\gamma (\mathbf{b})\mathbb{A}_{3}^{\prime +}+\delta (%
\mathbf{b})\mathbb{A}_{4}^{\prime +}\}|\Phi ^{\prime }),  \nonumber \\
\mathbb{U}(\mathbf{a},\mathbf{b})\mathbb{A}_{2}^{+}\mathbb{A}_{3}^{+}|\Phi )
&=&\{\gamma (\mathbf{a})\mathbb{A}_{1}^{\prime +}+\delta (\mathbf{a})\mathbb{%
A}_{2}^{\prime +}\}\{\alpha (\mathbf{b})\mathbb{A}_{3}^{\prime +}+\beta (%
\mathbf{b})\mathbb{A}_{4}^{\prime +}\}|\Phi ^{\prime }).
\end{eqnarray}%
Hence the final state $|\Psi ^{\prime })$ on which Alice and Bob perform
their measurements is%
\begin{eqnarray}
|\Psi ^{\prime }) &=&\frac{1}{\sqrt{2}}\left\{ [\alpha (\mathbf{a})\gamma (%
\mathbf{b})-\gamma (\mathbf{a})\alpha (\mathbf{b})]\mathbb{A}_{1}^{\prime +}%
\mathbb{A}_{3}^{\prime +}+[\alpha (\mathbf{a})\delta (\mathbf{b})-\gamma (%
\mathbf{a})\beta (\mathbf{b})]\mathbb{A}_{1}^{\prime +}\mathbb{A}%
_{4}^{\prime +}+\right.  \nonumber \\
&&\left. \lbrack \beta (\mathbf{a})\gamma (\mathbf{b})-\delta (\mathbf{a}%
)\alpha (\mathbf{b})]\mathbb{A}_{2}^{\prime +}\mathbb{A}_{3}^{\prime
+}+[\beta (\mathbf{a})\delta (\mathbf{b})-\delta (\mathbf{a})\beta (\mathbf{b%
})]\mathbb{A}_{2}^{\prime +}\mathbb{A}_{4}^{\prime +}\right\} |\Phi ^{\prime
})  \nonumber \\
&&  \label{888}
\end{eqnarray}%
We are going to focus on one particular partial observation, $P(+\mathbf{a},+%
\mathbf{b}|\Psi )$, which asks for the probability that Alice observes a
signal in ESD$_{1}$ \textbf{and} Bob observes a signal in ESD$_{3}$. In SQM
this corresponds to each observer catching their respective constituent
particle in its up state. From (\ref{888}) we immediately read off the
required amplitude, giving the probability%
\begin{equation}
P(+\mathbf{a},+\mathbf{b}|\Psi )\equiv (\Psi ^{\prime }|\mathbb{P}%
_{1}^{\prime }\mathbb{P}_{3}^{\prime }|\Psi ^{\prime })=\frac{1}{2}|\alpha (%
\mathbf{a})\gamma (\mathbf{b})-\gamma (\mathbf{a})\alpha (\mathbf{b})|^{2}.
\end{equation}

Wigner gave an intuitive calculation of a Bell-type inequality for such
observations \cite{WIGNER-1972}, arriving at the classical result%
\begin{equation}
P(+\mathbf{a},+\mathbf{b}|\Psi )+P(+\mathbf{b},+\mathbf{c}|\Psi )\geqslant
P(+\mathbf{a},+\mathbf{c}|\Psi ),  \label{666}
\end{equation}%
for any choice of three-vectors $\mathbf{a},\mathbf{b}$ and $\mathbf{c}$. In
our terms, this means that the coefficients have to satisfy the constraint%
\begin{equation}
|\alpha (\mathbf{a})\gamma (\mathbf{b})-\gamma (\mathbf{a})\alpha (\mathbf{b}%
)|^{2}+|\alpha (\mathbf{b})\gamma (\mathbf{c})-\gamma (\mathbf{b})\alpha (%
\mathbf{c})|^{2}\geqslant |\alpha (\mathbf{a})\gamma (\mathbf{c})-\gamma (%
\mathbf{a})\alpha (\mathbf{c})|^{2},
\end{equation}%
in addition to the semi-unitarity conditions already in force.\ It is easy
to find coefficients which violate this inequality. For example, following
Wigner, we take $\alpha (\mathbf{a})=\cos (\frac{1}{2}\theta _{\mathbf{a}})$%
, $\beta (\mathbf{a})=\sin (\frac{1}{2}\theta _{\mathbf{a}})$, $\gamma (%
\mathbf{a})=-\sin (\frac{1}{2}\theta _{\mathbf{a}})$ and $\delta (\mathbf{a}%
)=\cos (\frac{1}{2}\theta _{\mathbf{a}})$, where $\theta _{\mathbf{a}}$ is
real, and similarly for the other two rotations. Then each set of rotation
coefficients satisfies the semi-unitarity conditions and Wigner's inequality
reduces to%
\begin{equation}
\sin ^{2}(\theta _{\mathbf{a}}-\theta _{\mathbf{b}})+\sin ^{2}(\theta _{%
\mathbf{b}}-\theta _{\mathbf{c}})\geqslant \sin ^{2}(\theta _{\mathbf{a}%
}-\theta _{\mathbf{c}}).
\end{equation}%
It is easy to find three angles for which this condition is violated, which
demonstrates that QM is inconsistent with the sort of classical realism
which led to the Bell inequality (\ref{666}).

This result is consistent with Einstein locality because the partial
observation used involves both Alice and Bob together, i.e., treats both as
simultaneously local. On the other hand, partial observations involving $%
\mathcal{Q}^{1}$ and $\mathcal{Q}^{2}$ alone (i.e., by Alice alone), or
involving $\mathcal{Q}^{3}$ and $\mathcal{Q}^{4}$ alone (i.e., by Bob
alone), would be completely unaffected by whatever the other observer had
done to the axis of their particular S-G apparatus.

\section{Concluding remarks}

The above results fully support the position taken by Heisenberg and Bohr:
it is the experimental context alone which affects quantum outcome
probabilities, both in the preparation of labstates and in how they are
observed. Everything else is metaphysical speculation.

By showing that it is really the relationship between observers and
apparatus rather than SUOs that matters in quantum physics, these results
suggest that the status of quantum mechanics should be changed in a rather
serious way. Instead of physical reality being regarded as some
\textquotedblleft quantized\textquotedblright\ version of a classical
reality, quantum mechanics should be seen as no more and no less than the
correct and universal set of rules for information exchange between
observers and apparatus.

There are implications of this conclusion for various theoretical
disciplines such as quantum gravity and quantum cosmology. In those fields,
conventional approaches to quantization start by regarding space and/or the
universe as some sort of quantized SUO. The Bohr-Heisenberg vision of
reality, supported by QDN, suggests that those fields are ultimately doomed
to failure \emph{as they are currently formulated}, because quantum
mechanics cannot be discussed properly without sensible notions of observers
and apparatus, and such things could not have existed in proposed early
universe scenarios.

\end{document}